\documentclass[pra,showpacs,twocolumn,showkeys]{revtex4-1}
\usepackage{graphics}
\usepackage{amsmath}
\usepackage{mathtools}
\usepackage{color}
\usepackage{bbold}
\usepackage[usenames,dvipsnames]{xcolor}
\usepackage{dsfont}
\usepackage{braket}
\usepackage[framemethod=tikz]{mdframed}
\usepackage{cancel}
\usepackage{fancyheadings}
\begin{document}
\newcommand{\bstfile}{osa}
\newcommand{\bibs}{d:/Dropbox/Dad/Mark/References/BibFile}
\title{Length as a Paradigm for Understanding the Classical Limit}
\author{Mark G. Kuzyk}
\affiliation{Department of Physics and Astronomy, Washington State University, Pullman, Washington  99164-2814 \\ \today}

\begin{abstract}
It is regrettable that the quantum length of an object is rarely if ever discussed, because it provides an ideal pedagogical paradigm for understanding how a physicist uses classical intuition to define quantum properties and how such quantum properties behave as one would expect in the classical limit.  It also provides for a way to understand many-particle states, and leads to interesting quantum behavior that challenges our intuition of measurement.  This exercise parallels the ways that theories are developed, giving the student a concrete example of the thought process involved.
\end{abstract}

\maketitle

\section{Introduction}

The measurement of a length requires a comparison between a standard and the extreme ends of the object.  This view implicitly assumes that the object has sharp boundaries.  In the small size limit, where quantum effects dominate, the measurement is intrinsically fuzzy because the boundaries of both the ruler and the object are ill-defined, in which case the definition of length must be generalized.  For a small-enough object, the fuzziness of the object and ruler become comparable, challenging the concept of a classical length.

This paper argues for a definition of length in the classical limit using an intuitive approach, applies it to the quantum realm and shows how the definition asymptotes to the expected classical behavior in the many-particle limit by building a material from units cells.  This approach is applied to both an object and a ruler to investigate the essence of a measurement, then treats the ruler and object together as a quantum system, where all electrons need to be entangled.  The classical limit gives the expected results but the quantum limit under certain special cases leads to strange phenomena.

The approach presented here in developing a formalism for the quantum length parallels the way theories are more generally developed by physicists.  The principles can be summarized by four steps.  {\bf The physicist}:
\begin{enumerate}

\item {\bf identifies the physical constraints.}  Section \ref{sec:ClassicalLength} argues that our notion of length requires it to be translationally invariant and must give the correct classical value.

\item {\bf chooses the simplest ansatz that meets the constraints.}  Section \ref{sec:ClassicalLength} applies translational invariance to propose the simplest ansatz for the classical length of a linear and uniform rod, which leads to an expression that has the same form as the position uncertainty in the quantum realm.  Appendix \ref{sec:NonuniformRod} shows that the result applies to a nonuniform classical rod and Appendix \ref{sec:Third-Order-Length} proposes a more complex ansatz for the length to illustrate that many theories might fit the broad requirements of a theory.

\item {\bf proposes a theory that holds in a new realm and demands that it agrees in the classical limit.}  Section \ref{sec:QuantumLengthCalc} proposes the theory of the quantum length as the position uncertainty even when the probablity density is non-uniform.  Section \ref{sec:applications} derives the wave functions of $N$ noninteracting fermions and shows that the classical result is obtained for large $N$ and for one particle in the limit of it occupying the highest-energy state of the system.

\item {\bf investigates the consequences.} Section \ref{sec:Bosons} applies the theory to bosons and Section \ref{sec:Rulers} applies it to rulers and length measurement to illustrate how the theory is nonintuitive in the quantum realm.  This discussion leads to an appreciation for the difference between quantum and classical parsing which leads to Section \ref{sec:Learned} showing that beyond a critical resolution the accuracy of a ruler does not get better by making finer markings.

\end{enumerate}

\section{Classical Length}\label{sec:ClassicalLength}

\begin{figure}[h]
\vskip 0in
\hspace{-.5em}\includegraphics{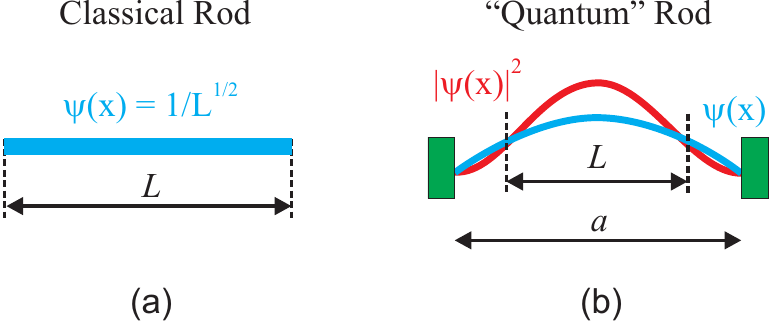}
\caption{(a) A classical rod can be represented by a uniform {\em material} density with sharp edges. (b) The ground state wave function of an electron in a box has a {\em probability} density that is proportional to the square of the wave function.}
\label{fig:1D-Box}
\end{figure}

Classical objects have an approximately uniform density and sharp boundaries.  To make a connection with a quantum object, we first take an inventory of the types of measurables associated with spacial properties, including:  (1) the expectation value of the position -- which gives the average position, and the variance -- which quantifies the spread of the wave function.\cite{sakur10.01,cohen77.01}  We use these observables to propose a quantum-like definition using a classical rod of length $L$ as shown in Fig.~\ref{fig:1D-Box}a.

For a uniform rod, we can define a classical linear density amplitude
\begin{equation}\label{eq:ClassicalPsi}
\psi(x) =  \left\{
\begin{array}{l l}
 \frac {1} {\sqrt{L}}, & \mbox{ for } 0 \le x \le L \\
 0, & \mbox{ for } x < 0 \mbox{ and } X>L \\
 \end{array}
                               \right.
,
\end{equation}
which is normalized, so $\int_{-\infty}^{+ \infty} \left| \psi(x) \right|^2 \, dx = 1$.  $\left| \psi(x) \right|^2 $ is the normalized density of material at point $x$.  The density is constant in the rod and vanishes outside of it.  When multiplied by the mass of the rod it gives the mass density, and when multiplied by the number of particles gives the particle number density, etc.

Uniformity is a classical notion that the material is the same at any point within it.  The formalism presented here can be extended to a nonuniform rod by viewing it as a series of classically uniform segments as described in Appendix \ref{sec:NonuniformRod} For simplicity, we will focus on the uniform classical rod as an element of length.

The expectation of the position for the density amplitude given by Equation \ref{eq:ClassicalPsi} is
\begin{equation}\label{eq:<x>}
\left< x \right> = \int_0^L x \left| \psi(x) \right|^2 = \frac {L} {2}
\end{equation}
and $\left<x^2 \right>$ is given by
\begin{equation}\label{eq:<x^2>}
\left< x^2 \right> = \int_0^L x^2 \left| \psi(x) \right|^2 = \frac {L^2} {3} .
\end{equation}
Equations \ref{eq:<x>} and \ref{eq:<x^2>} are not solely the properties of the rod, as can be simply verified by shifting the rod by a distance $\delta$, yielding
\begin{equation}\label{eq:<x>shift}
\left< x \right> = \int_{\delta} ^{L+ \delta} x \left| \psi(x) \right|^2 = \frac {L} {2} + \delta
\end{equation}
and $\left<x^2 \right>$ is given by
\begin{equation}\label{eq:<x^2>shift}
\left< x^2 \right> =\int_{\delta}^{L+ \delta} x^2 \left| \psi(x) \right|^2 = \frac {L^2} {3} + L \delta + \delta^2.
\end{equation}

However, $\Braket{x^2} - \Braket{x}^2$ is independent of the position $\delta$, as we can see using Equation \ref{eq:<x>shift} and  \ref{eq:<x^2>shift}, yielding for the uniform rod
\begin{equation}\label{eq:UncertainSquare}
\Braket{x^2} - \Braket{x}^2 = \frac {L^2} {12}.
\end{equation}
This is just the position uncertainty $\Delta x = \sqrt{ \left< x^2 \right> - \left< x \right>^2}$, which for the uniform rod yields
\begin{equation}\label{eq:DeltaX}
\Delta x = \sqrt{ \left< x^2 \right> - \left< x \right>^2} = \frac {L} {\sqrt{12}}.
\end{equation}
Since Equation \ref{eq:DeltaX} is the simplest quantity with units of length that is independent of the position of the rod, we use Equation \ref{eq:DeltaX} as the basis for the ansatz that the length is given by
\begin{equation}\label{eq:ClassicalLength}
L = \sqrt{12} \Delta x = \sqrt{12 \left( \Braket{x^2} - \Braket{x}^2 \right)}.
\end{equation}

If the object is not uniform, the classical length is obtained by summing over the infinitesimal lengths of its uniform parts as described in Appendix \ref{sec:NonuniformRod}, leading to Equation \ref{eq:LengthAgnosticIntegrate} for the more general form.

Equation \ref{eq:ClassicalLength} is derived by choosing the simplest expression that is translationally invariant and gives the correct length for a classical rod.  More complex expressions that fulfill the requirements are possible, but they are not the simplest expressions for the length.  Appendix \ref{sec:Third-Order-Length} shows an example where expectation values contain terms to third power in the position.

The wave function replaces the material density amplitude in the quantum realm.  Thus, the interpretation also changes; rather then representing the material density amplitude, the wave function $\psi(x)$ represents the probability amplitude that a measurement will find material at point $x$.  Electromagnetic-based measurements such as light scattering -- which we use to ``see" a material, and contact forces -- which we use to ``feel" a material probe a material's electrons.  Thus in the classical limit, where a chunk of material contains many electrons, the two definitions converge to the same value; that is, the number of electrons that fill a volume (i.e. the density of the material) is also proportional to the probability of observing them.

We can express the length along $x$ in agnostic form that works equally well for the classical and quantum case, given by
\begin{equation}\label{eq:LengthAgnostic}
\boxed{  L = \sqrt{12  \int \rho( x ) \left(  x^2 - \bar{x}^2 \right) dx, }}
\end{equation}
where $\rho(x)$ is the dimensionless density with $\int dx \, \rho(x) = 1$ and $\bar{x}$ is the $x$ coordinate of the ``center" of the object.  The integral is over all space.  In this form, $\rho( x)$ can be the classical dimensionless density or the quantum probability density $\psi^*( x ) \psi( x )$.  The length in any direction can be calculated by rotating the object so that the coordinate x-axis is along the desired dimension of measurement.  This is no different than laying a ruler across an object along the desired direction.  Equation \ref{eq:LengthAgnostic} is the central definition that is the subject of the investigations that follow.

We later test this definition by checking that the quantum version of the rod asymptotes to the expected length in the classical limit when the number of particles (i.e. electrons) making the object becomes large or the state energy tends to infinity for a single particle.

\section{Quantum Length}\label{sec:QuantumLengthCalc}

This section develops the calculation of the quantum length, which cannot be calculated in the classical way of breaking the material into small sections and adding their lengths.  The particle in a box illustrates why.

In the ground state, the probability density peaks in the middle of the well, as shown in Fig.~\ref{fig:1D-Box}b.  If the box is split in half adiabatically, the wave function will be peaked in each of the smaller boxes, so the probability density changes.  In contrast, the classical material density remains the same when a piece of the material is split in half, making additivity possible as quantified by Equation \ref{eq:LengthAgnosticIntegrate}.  At the heart of the problem in the quantum view is that the act of splitting up the material into pieces changes the material.  Classically, the material remains the same.

Secondly, there is a difference between the meaning of probability density given by the wave function and the material density.  The probability density at a given position is defined as the probability per unit volume of finding material (let's say an electron) there.  The classical material density, on the other hand, can be used to determine the amount of material in a given volume.

Equating the quantum probability density with the classical density poses issues for generalizing the classical definition to the quantum realm.  In the classical view, an object has a boundary beyond which no material of the object is found.  The quantum probability, on the other hand, can be nonzero everywhere -- as it would be for an non-infinite box, making the quantum length infinite.

We can reconcile these two extremes by noting that the classical material can be viewed as being made of so many quantum boxes that subdividing a material amounts to separating it into smaller spatial groupings of boxes with still a vast number of boxes in each grouping.  Then, no individual quantum boxes are being divided.  The assumption we will make here, which will later be tested, is that the determination of the length of a quantum box is through Eq.~\ref{eq:LengthAgnostic} even when the wave function is not uniform.  Eq.~\ref{eq:LengthAgnosticIntegrate} will hold for a collection of boxes that are so large in number that an ``infinitesimal" piece of material contains many quantum boxes so that the element is uniform on macroscopic scales.  

Throughout the rest of the paper, we assume that electrons are the particles that mediate classical interactions.  This section calculates the length for a single-electron state then for a many electron state using the fact that each electron in a material, being identical to all others, is spread over the whole material.

\subsection{Single-Particle Length}\label{sec:SingleParticleLength}

For one electron in a box, the ground state wave function
 \begin{equation}\label{eq:GroundStateBoxPsi}
\psi(x) = \sqrt{ \frac {2} {a}} \sin \left( \frac {\pi x} {a} \right)
\end{equation}
is plotted in blue in Fig.~\ref{fig:1D-Box}b.  The square of the wave function is shown in red.  Is the size of the material object made of the elecron in the box the distance between the walls?  If so, half the wavelength of the ground state wave function defines the size, which is given by $a$.

To see why this is not so, consider the fact that contact forces between objects are what allow the dimensions of a wood block to be measured with a vernier calliper; and, these forces originate in the electrons.  The nuclei provide the scaffolding for the electrons and the electrons in one object interact with the electrons in the another one.  So when you touch a table, the electrons in your finger repel the electrons in the table.  To carry forward this analogy, the particle in the box is an electron and the box provides the confinement potential as do nucleons in atoms and molecules.  When we ``touch" the box, we are in fact touching the electron cloud.  The walls of the box merely provide the force that keeps the electrons confined and is not a ``material" object that we can sense no more than we can directly sense the nucleons in everyday interactions.

Models of materials with non-interacting electrons in a box roughly predict the electronic properties of small molecules such as the molecular class of polyenes,\cite{kuhn48.01} describe metals as found in solid state textbooks\cite{OpenStax20.01} and accurately portray the quantum to classical transition of nano-particles.\cite{schol12.01}  This shows that the effect of the nucleons on the electrons can be roughly taken into account with a box that confines the electrons within.  We will thus model typical materials with uniform electron density as non-interacting electrons in a box.  The reader should keep in mind that this is a first step in modelling materials in which electrons are delocalized.  Later we treat materials made of such units that are ``pressed" together.  Then, the electrons are localized within domains rather than over the full material.  For simplicity, we will treat only one-dimensional systems.  Other potentials can be treated in the same way, but this exercise does not result in significant-enough insights about length itself to make it worthwhile to treat in this paper.

The quantum size of the system, we argue, is determined from the breadth the square of the wave function as shown in red in Fig.~\ref{fig:1D-Box}.  The uncertainty in its position is given by
\begin{equation}\label{eq:DeltaX-QuantumBox}
\Delta x = a \sqrt{ \frac {1} {12} - \frac {1} {2 \pi^2}} .
\end{equation}
Substituting Equation \ref{eq:DeltaX-QuantumBox} into Equation \ref{eq:ClassicalLength} yields the length of a quantum particle in a box, or
\begin{equation}\label{eq:Length-QuantumBox}
L = a \sqrt{ 1 - \frac {6} {\pi^2}} \approx 0.63 a.
\end{equation}
The length given by Equation \ref{eq:Length-QuantumBox} defines an interval over which the probability density remains above about 30\% of its peak value.  This picture is a fully quantum one that offends our common sense that objects have well-defined boundaries, but the ansatz given by Equation \ref{eq:ClassicalLength} provides a definition of this object's length.

\subsection{Other Invariants}

\subsection{Many-Particle Length Using Single-Particle Operators}

The reader is encouraged to work through the appendices, which introduce the specialized conventions used here and to review how dirac notation is used in many-particle systems.  Below, we calculate the quantum length of many non-interacting electrons as described by Equation \ref{slater} using the fact that each electron spreads out over the full length of the object -- and because they are indistinguishable -- the spread of any single electron will be representative of the whole many-electron object.

We choose to get the length from Particle \#1, whose position operator is given by $x_1$, which in the system's ground state $\Ket{G}$ has a spread
\begin{equation}\label{LengthFromOne}
\Delta x_1^2 = \Braket{G|x_1^2|G} - \Braket{G|x_1|G}^2 .
\end{equation}
Using Equation \ref{ExpectManyFinal} with $n_i = i$, Equation \ref{LengthFromOne} takes the form
\begin{align}\label{SpreadManyFinal}
\Delta x_1^2 &= \frac {1} {N} \sum_{n=1}^N \Big( \Braket{n|x^2|n}  \nonumber \\
& - \frac {1} {N} \Braket{n|x|n} \sum_{n^\prime = 1}^N  \Braket{n^\prime|x|n^\prime}\Big),
\end{align}
where we have dropped the subscript identifying Particle \#1 since this result holds for any particle.  Using Equation \ref{eq:ClassicalLength}, the length is then given by
\begin{align}\label{DesnityNFinal2}
L &= \bigg( \frac {12} {N} \sum_{n=1}^N  \Big( \Braket{n|x^2|n}  \nonumber \\
& - \frac {1} {N} \Braket{n|x|n} \sum_{n^\prime = 1}^N  \Braket{n^\prime|x|n^\prime}\Big) \bigg)^{1/2} .
\end{align}

When the potential is centrosymmetric about position $x=0$, the single-particle energy eigenfunctions have a definite parity, i.e.
 \begin{equation}\label{eq:Centrosymmetry}
\psi_n (-x_i) = \pm \psi_n (x_i) ,
\end{equation}
so
\begin{equation}\label{eq:CentrosymmetryPosition}
\Braket{n|x_i|n} = \int_{-\infty}^{+\infty} dx_i \, \psi^*(x_i) x_i \psi(x_i) = 0 .
\end{equation}
If the center of symmetry is shifted by $x \rightarrow x + c$, then
\begin{align}\label{eq:Shift Center C}
\Braket{n|x_i|n} = c
\end{align}
for all $n$.  With the help of Equation \ref{eq:Shift Center C}, Equation \ref{DesnityNFinal2} can be expressed as
\begin{equation}\label{eq:SimplifiedL-Final}
\boxed{
\begin{aligned}
& L =  2 \sqrt{3}  \left( \frac {1} {N} \sum_{n=1}^{N} \left<  n \right| x_i^2  \left|  n \right> - c^2 \right)^{1/2} . \\
& \mbox{Many-Particle Wave function} \\
& \mbox{Centrosymmetric Potential}
\end{aligned}
}
\end{equation}
The object's length given by Equation \ref{eq:SimplifiedL-Final} is expressed in terms of single-particle states and operators.

\section{Building a Material}\label{sec:applications}

Here we apply the above formalism with non-interacting electrons to calculate the length of a particle in a box and evaluate the classical asymptote when the number of electrons becomes large.  We also plot the electrons densities to determine if the asymptote meets our classical expectations and interpret the results in terms of classical parameters.

\subsection{Particle in a Box}

The particle in a box is the ideal system for studying length by virtue of its sharp walls.  We have two choices for building a box.  First, we can fix the box's size, then fill it with electrons.  A more realistic model is to increase the size of the box in proportion to the number of electrons, which more closely reflects a real system such as a metal.  With each added atom, the lattice become larger and simultaneously gains conduction electrons.  We calculate both cases, starting with the simpler case of the fixed box size.

\begin{figure}[h]
\vskip 0in
\hspace{0em}\includegraphics{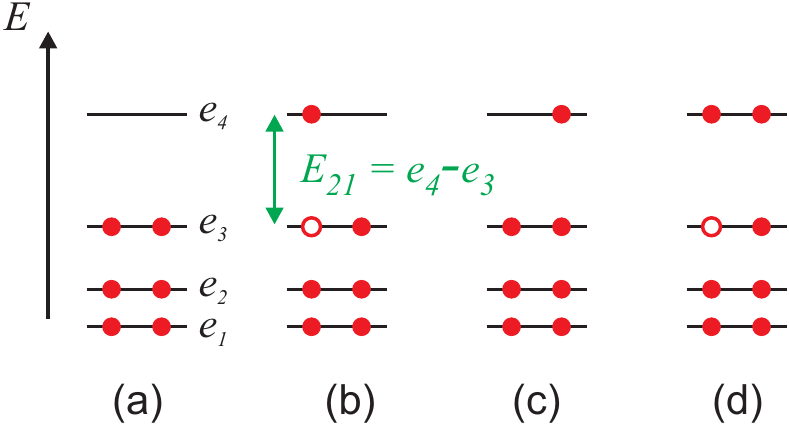}
\caption{(a) The ground state and (b) the first excited state of $N$ particles in a box when $N$ is even; and (c) ground state and (d) first excited state when $N$ is odd.}
\label{fig:FixedBox}
\end{figure}

The quantum calculations of the length of a particle in a box starts with determining $\Braket{n|x^2|n}$.  For convenience, we place the left-hand side of the box at $x=0$ and make its length $a$, so
\begin{equation}\label{eq:x^2Box}
\Braket{n|x^2|n} = \frac {2} {a} \int_0^a dx \, x^2 \sin^2 \left( \frac {n \pi x} {a} \right) .
\end{equation}

Equation \ref{eq:x^2Box} can be integrated by making the trigonometric substitution $\sin^2 \theta = (1 - \cos 2\theta)/2$; integrating the second term by parts, twice; and then using the fact that the sine function vanishes at the endpoints, yielding
\begin{equation}\label{eq:x^2BoxMatrix}
\Braket{n|x^2|n} = a^2 \left( \frac {1} {3} - \frac {1} {2 \pi^2 n^2} \right).
\end{equation}

Now we evaluate the length of the ground state of a system that has two electrons in each of the lowest-energy states -- one spin up and one spin down. Substituting Equation \ref{eq:x^2BoxMatrix} into Equation \ref{eq:SimplifiedL-Final} and using the fact that the square well is a centrosymmetric potential centered at $c = a/2$ yields the $N-$particle length
\begin{equation}\label{eq:PIB-Length}
\boxed{
\begin{aligned}
& L_N^2 =   a^2 \left(   1 - \frac {12} {\pi^2 N} {\sum_{n=1}^{N/2}}^\prime \frac {1} { n^2} \right). \\
& \mbox{ Length of N-particles in a box}
\end{aligned}
}
\end{equation}
In making the above substitution, we use the fact that the summation in Equation \ref{eq:SimplifiedL-Final} is over all occupied states.  A spin up or spin down electron placed in the same energy eigenstate is considered distinct.  Equation \ref{eq:PIB-Length}, on the other hand, sums over only the occupied energy eigenstates.  When electrons do not interact, the spin-up and spin-down electrons give the same length.  The restricted sum is signified by the prime, with $\sum_{n=1}^{N} \rightarrow 2{\sum_{n=1}^{N/2}}^\prime$, that is, the highest occupied state is $n = N/2$ with two electrons per state.    Equation \ref{eq:PIB-Length} in the two-electron limit, with $N=2$ agrees with Equation \ref{eq:Length-QuantumBox}.

In the classical limit with many electrons, Equation \ref{eq:PIB-Length} gives
\begin{equation}\label{eq:PIB-Length-Clasical}
\lim_{N \rightarrow \infty} L_N =  \lim_{N \rightarrow \infty} a \sqrt {\left( 1 - \frac {2} {N} \right)} = a ,
\end{equation}
where we have used the fact that $\sum_{n=1}^\infty n^{-2} = \pi^2 / 6$. Equation \ref{eq:PIB-Length-Clasical} makes intuitive sense: the electron cloud's length is the distance between the walls of the box, which define the classical turning points.

If instead we build a system by adding unit cells of length $a_0$ each with two electrons,\cite{kuzyk16.01} we can treat this case simply by making the substitution $a =  a_0 N/2$ in Equation \ref{eq:PIB-Length}, then the length of the box for $N \rightarrow \infty$ asymptotically approaches $N a_0/2$.  Again, the asymptote gives the distance between the classical turning points.

Alternatively, we can use the above analysis to determine the length of a single electron in a box in which only state $\Ket{n}$ is occupied by including only this one term in the sum with $N=2$ in Equation \ref{eq:PIB-Length}, yielding
\begin{equation}\label{eq:BoxExcite-state-length}
\boxed{
\begin{aligned}
& L_n = \sqrt{12 \left( \Braket{n|x^2|n} - \frac {a^2} {4} \right)} = a \sqrt{ 1 - \frac {6} { \pi^2 n^2} } . \\
& \mbox{Excited State Length of Single Particle  } \\
& \mbox{in a Box in State $\Ket{n}$}
\end{aligned}
}
\end{equation}
The excited-state length defined by Equation \ref{eq:BoxExcite-state-length} approaches the length of the box in the limit of large energy ($\lim_{n \rightarrow \infty} L_n = a$).  Thus, the length of the system defined by the electron cloud approaches the distance between the walls of the box when either the number of electrons becomes large or when one electron populates a high-energy state.

\subsection{Electron Density}

The electron density provides a useful visualization of the classical limit because it is what we interact with.

\begin{figure}[h]
\vskip 0in
\hspace{-.5em}\includegraphics{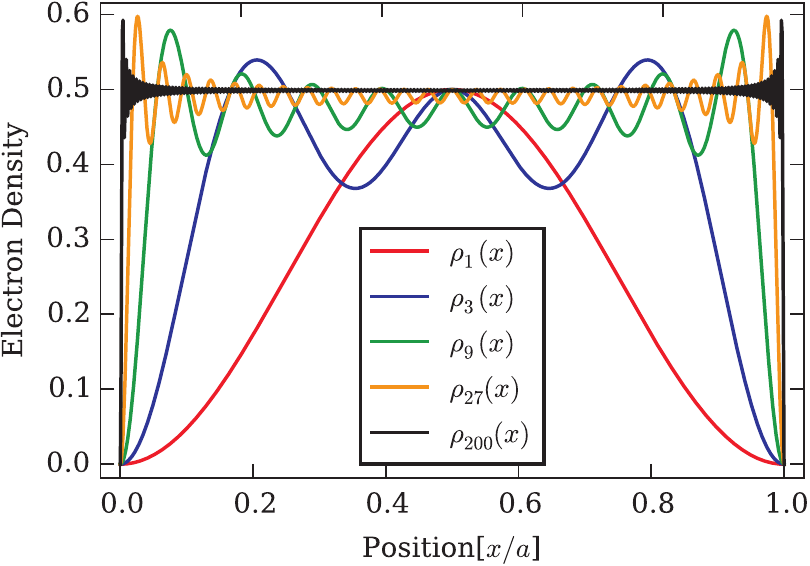}
\caption{The electron density in a box as a function of the highest occupied state $n=$1, 3, 9, 27, and 200.  In the classical limit, where the number of electrons becomes large, a uniform probability density is observed.}
\label{fig:ParticleDensity}
\end{figure}
The probability density for one of the electrons given by $\rho(x) = \left| \psi(x) \right|^2 = \left| \Braket{x_1|G} \right|_{x_1 = x}^2 $ can be simply calculated from Equation \ref{eq:GeneralState}, giving
\begin{equation}\label{eq:OneParticleDenisty}
\rho(x) = \left| \psi(x) \right|^2 = \frac {1} {N} \sum_{n=1}^{N} \left| \psi_n (x) \right|^2 .
\end{equation}
Figure \ref{fig:ParticleDensity} plots the electron density as a function of position within a box of width $a$.  Shown are electron densities for 2, 6, 18, 54, and 400 electrons; which fill states up to 1,3, 9, 27, 200 (2 electrons per state).  This plot makes clear how the density becomes uniform with sharper edges in the classical limit when $N \rightarrow \infty$, an effect observed in nano-spheres.\cite{schol12.01}

Note that Equation \ref{eq:OneParticleDenisty} is reminiscent of a mixed state, so can be used in pedagogy surrounding the mysterious density matrix,\cite{landa69.01} a topic that will not be discussed here.

\section{Bosons}\label{sec:Bosons}

It is interesting to think about how the world would appear if the constituents of matter were bosons.  Let's next consider a system made of noninteracting bosons.  Since the state vector is symmetric upon interchange of any two particles, the ground state is the one where all $N$ particles occupy the single-particle ground state, or
\begin{equation}\label{eq:BosonWavefucntion}
\Ket{G} = \Ket{1,1, \dots 1}.
\end{equation}

The length of a system of non-interacting bosons in a box is simple to compute using the single-particle method with the wave function given by Equation \ref{eq:BosonWavefucntion}, yielding
\begin{equation}\label{eq:PIB-BosonLength}
L_{box} =   a \sqrt{  1 - \frac {6} { \pi^2 } } ,
\end{equation}
which is independent of the number of bosons.   In contrast, the length of one fermion in a box is the same as for a boson, but the length increases as fermions are added, converging to the distance between the walls.  Not so for bosons.

The density of Particle 1 of the ground state of a collection of Bosons is given by
\begin{equation}\label{eq:BosonDensity}
\rho (x_1) =  \left|\Braket{x_1 | G} \right|^2 = \left| \Braket{x_1 | 1,1, \dots 1} \right|^2 = \left| \psi_1 (x_1) \right|^2.
\end{equation}
Thus, it is not surprising that the quantum length of the many-particle system of bosons in their ground state is the same as the single-particle result for any fermion system in its ground state.

We saw for fermions that the particle density function in the large-particle-number limit approaches a step function for a particle in a box.  For bosons, the particle densities for the box are peaked functions, and the shape is independent of the number of particles.  This behavior contradicts our intuition of how a system should behave because the materials with which we interact are made of fermions.  In a world of bosons, rulers would not have sharp ends so length would not have developed as an obvious concept.

Note that we could in principle construct a boson state vector that is of the same form as the fermion state vector given by Equation \ref{slater2} except that all the signs are positive for bosons.  Then, it is easy to show that the boson length is the same as for fermions since the signs do not change the result.  Then, the bosonic system would behave as we expect in the classical limit, making rulers with sharp boundaries.  So, bosons can be prepared in an excited state that makes their classical limit appear fermion-like; but fermions cannot be prepared in a state that mimics bosons because of Pauli exclusion.

\section{Discussion}\label{sec:Rulers}

Measurement is central to physics.  Since we are interested in length, this section focuses on the use of a ruler, the ramifications of making a subdivision and its accuracy.  We also address the implications of the requirement that a ruler and the object being measured must be entangled since they are both made of fermions.  A boson ruler is also discussed.

\subsection{Rulers}

\begin{figure}[h]

\includegraphics[width=3.3in]{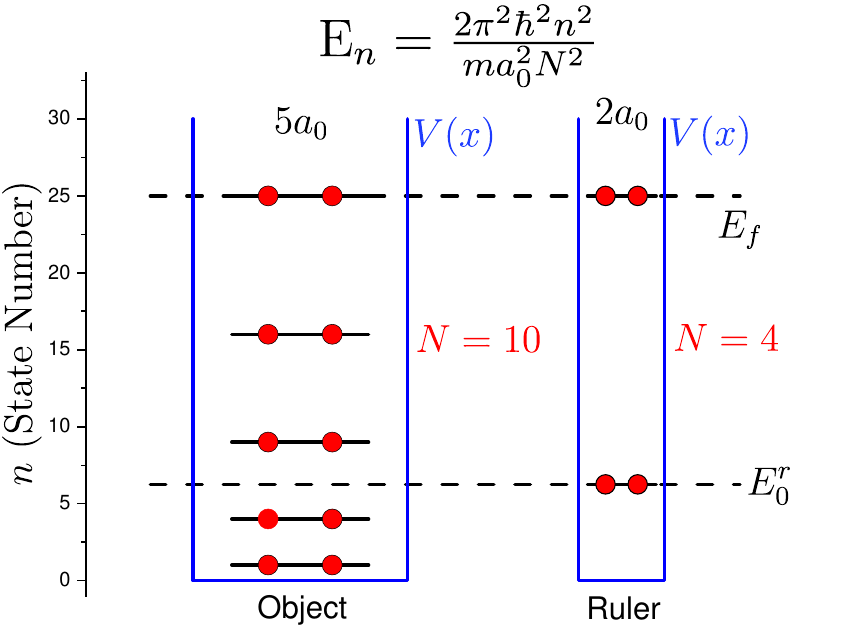}
\caption{An object's length is determined using a ruler made of a line of small identical pieces or a single piece with markings.  Shown here is an object and a ruler segment as an entangled quantum system.  $N$ is the number of electrons with two electrons per unit cell.}
\label{fig:ObjectRuler}
\end{figure}
The ancients used a cubit, the distance between the elbow and tip of the middle finger, as a convenient unit of measurement.\cite{robin07.01} The builders would successively place their arms along an object and count the number of cubits spanned by the object.  Such measurements require that the object and arm be well-enough defined to locate their edges.

Rather than making lots of small identical rods, a ruler is a single object that is marked at equal intervals, which can be viewed as a stack of many tiny rods of equal length.  If the markings are to be resolvable, the de Broglie wavelength of the electrons in the ruler material must be shorter than the width of the markings.   A ruler with markings obeys the same quantum physics as a stack of non-interacting rods because two consecutive marks localizes that segment.  We thus use a stack of rods as the physical picture of a ruler, each rod represented by an electron in a box with a length defined by Equation \ref{eq:LengthAgnostic}.

Figure \ref{fig:ObjectRuler} shows an object made of 5 unit cells and one ruler segment made of two unit cells.   The object and ruler are assumed to be made of the same material, so the Fermi energies are the same.\cite{ashcr76.01}  The length of the ruler is defined by the sum of the lengths of its quantum segments when they are brought together.  If a uniform rod is placed next to the object and observed to be of the same length, our classical world finds that placing markings on the ruler changes nothing.  However, in the quantum realm, the ruler will become shorter when markings are added because the electron cloud will occupy an ever smaller part of the box defined by the distance between markings, approaching a length of 0.63 $a_0$ in the limit of two electrons per segment of unit cell size, as we saw in Equation \ref{eq:Length-QuantumBox}.  We quantify these observations as follows.

Consider a ruler of length $L$ that is determined from visual inspection to be of the same size as the object that is being measured.  We assume the ruler is made of $N/2$ unit cells (2 electrons per unit cell) each of length $a_0$ and containing two electrons, for a total of $N$ electrons.  The length of the $N$-electron box is given by the spread of the electron density.  In the limit when the number of electrons gets large, and the electrons are delocalized along the whole ruler, its length is given by $L = N a_0/2$, which coincides with the walls of the full box.  The length of a unit cell in isolation, which contains two electrons, is given by Equation \ref{eq:Length-QuantumBox}.

Next, we divide the rod into $R$ equal cells, where $1 \le R \le N/2$, to make a ruler where the contact points between the $R$
segments define the markings of the ruler.  We assume that the electrons are localized in their own segments; and, when the segments are pressed together, the length will be the sum of the quantum lengths of the segments.

Each segment, then, has $N/R$ electrons and the length of the box segment is $N a_0/2R$. We can get the length of one segment from Equation \ref{eq:PIB-Length} with the substitution $N \rightarrow N/R$ and $a \rightarrow N a_0/2R$, yielding
\begin{equation}\label{eq:PIB-Length-Cell}
\Delta L_R (N)= \frac {N a_0} {2R} \sqrt{1 - \frac {12} { \pi^2 } \cdot \frac {R} { N} \cdot {\sum_{n=1}^{N/2R}}^\prime} \frac {1} { n^2} ,
\end{equation}
where the highest state occupied is $n = N/2R$, so $n$ is an integer, meaning that a unit cell cannot be subdivided.  The length of the electron cloud, hence the ruler, is then given by $L_R(N) = R \Delta L_R (N)$, which using Equation \ref{eq:PIB-Length-Cell} yields
\begin{equation}\label{eq:PIB-Length-Add-Cells}
L_R(N) =  \frac {N a_0} {2}  \sqrt{1 - \frac {12} { \pi^2 } \cdot \frac {R} { N} \cdot {\sum_{n=1}^{N/2R}}^\prime \frac {1} { n^2}} .
\end{equation}
When $R=1$ and $N$ is large, Equation \ref{eq:PIB-Length-Add-Cells} gives the length $L_R = N a_0 / 2$, as we would expect for a rod with $N/2$ unit cells with fully delocalized electrons.

For ruler segments of unit cell size with two electrons in each, the ruler's length is
 \begin{equation}\label{eq:PIB-Length-Unit-Cells}
L_{R=N}(N) =  \frac {N a_0} {2}  \sqrt{ 1 - \frac {6} { \pi^2} } \approx 0.63 N a_0 / 2.
\end{equation}
Thus, chopping the ruler into the smallest possible segments and recombining them makes the ruler over a third shorter than its monolithic form.  Using the classical assumption that the length remains constant under such an operation would lead to an overestimate of the length of the object that is being measured.

An interesting artifact of chopping a rod into segments is that the process leads to a change in the energy of the rod.  The ground state energy of the rod with fully delocalized electrons is given by
\begin{align}\label{eq:GS-energy-FullRod}
E_G = 2 \sum_{n=1}^{N/2} \frac {h^2 n^2} {2 m N^2 a_0^2} = \frac {h^2 (N+2) (N+1) N} {24 m N^2 a_0^2}.
\end{align}
The ground state energy of a ruler sliced into $R$ segments is given by
\begin{align}\label{eq:GS-energy-Pieces}
E_G^\prime = 2R  \sum_{n=1}^{N/2R} \frac {h^2 R^2 n^2} {2 m N^2 a_0^2} = \frac {h^2 (N+2R) (N+R) N} {24 m N^2 a_0^2}.
\end{align}
The reader can easily verify that in the limiting case when the ruler is uncut so $R=1$, Equation \ref{eq:GS-energy-Pieces} is the same as Equation \ref{eq:GS-energy-FullRod}.  If the ruler is cut into its smallest possible parts, $R=N/2$ and $E_G^\prime / E_G =3$ when $N \rightarrow \infty$.  Thus, the net amount of work done in cutting up the ruler is twice the ground state energy of the monolithic rod.

\subsection{Ruler Reliability}

Next we consider a user of a ruler who assumes that its length is independent of the number of subdivisions, as would be the case within the realm of common experience.  As long as the classical limit is maintained, the precision of the measurement increases as more rulings are made.  Since the object being measured is large and behaves classically, its edges are sharp so can be located with almost arbitrary precision, to a level that we assume to be on the order of $L/R$, the spacing between markings.  However, as the rulings are made closer together and quantum effects take hold, the systematic overestimate of the object's length grows.

The length measurement is most true when the increase in resolution from making smaller segments is offset by a greater overestimate of the object's length -- the condition given by $Na_0/2 - L_R(N) = N a_0 /2R$.  Using Equation \ref{eq:PIB-Length-Add-Cells}, this condition gives
\begin{equation}\label{eq:PrecisionCondition}
1 -  \sqrt{1 - \frac {12} { \pi^2 } \cdot \frac {R} { N} \cdot {\sum_{n=1}^{N/2R}}^\prime \frac {1} { n^2}} = \frac {1} {R} .
\end{equation}
If the number of rulings $R$ is large when this condition is met, the second term under the square root in Equation \ref{eq:PrecisionCondition} is necessarily small.  Since the sum converges to $\pi^2 / 6$ for an infinite number of terms, it can be approximated by $ 2R/N  \ll 1$ for a finite but large number of terms.  Then, the square root can be expanded in a series with $R/N$ small, giving the condition
\begin{equation}\label{eq:IdealRule}
\boxed {
\begin{aligned}
&R = \sqrt{N} = \sqrt{2L/a_0}. \\
&\mbox{Ideal Number of Rulings} \\
& \mbox{for Maximum Accuracy} \\
& \Delta L = \frac {L} {R} = \sqrt{a_0 L/2} \\
& \mbox{Optimum Size of Rulings}
\end{aligned}
}
\end{equation}
Thus, for a ruler of 10 cm length made of a material with lattice constant $a_0 = 1 \mbox{\AA}$, the most accurate resolution is for $2 \, \mu m$ rulings.

\subsection{Classical Versus Quantum Parsing}

We have calculated the length of a single ruler and of the individual sections quantum mechanically.   However, when combining the sections, we are using the classical notion of the length, that is, the ruler length is taken to be the sum of the pieces.  This requires that two issues be resolved.  First, the assumption that the ruler segments can be stacked as if each piece is of the calculated quantum length leads to an overlap in the wave functions between them.  This is of no consequence if the particles, be they bosons or electrons, are noninteracting, as we have assumed from the start.  A second issue is that the classical construction ignores the requirement that the ruler segments are entangled.

To avoid undue complexity, we consider a ruler with $N$ particles and $N$ segments with one particle per segment.  Then, in the global ground state, each segment will also be found in its ground state.  If the single-particle ground state vector of the $i^{th}$ segment containing one particle is designated by $\Ket{1^{(i)}}$, the fully antisymmetric entangled state vector of the segmented ruler is given by
\begin{equation}\label{eq:EntangledSegments}
\Ket{G} = \frac {1} {\sqrt{N}} \sum_{i=1}^N \Ket{1^{(i)}} \Ket{\overline{1^{(i)}}} .
\end{equation}
The decomposition given by Equation \ref{eq:EntangledSegments} is analogous to Equation \ref{eq:GeneralState}, but rather than forming an entanglement of single particle states of different energies, the energies are the same and the entanglement is between the physically-separated segments.

Using Equation \ref{eq:EntangledSegments} to get the probability density yields
\begin{equation}\label{eq:EntangledDensity}
\rho(x_1) = \left| \frac {1} {\sqrt{N}} \sum_{i=1}^N \Braket{x_1 | {1^{(i)}} } \Ket{\overline{1^{(i)}}} \right|^2 =  \frac {1} {N} \sum_{i=1}^N \rho_1^{(i)}(x_1),
\end{equation}
where $\rho_1^{(i)}(x_1)$ is the probability density of the particle in segment $i$ in its ground state.  Note the similarity between Equation \ref{eq:EntangledDensity} and Equation \ref{eq:OneParticleDenisty}.  Equation \ref{eq:OneParticleDenisty} is the average density of $N$ particles in different energy eigenstates of the same box, but Equation \ref{eq:EntangledDensity} is the average density of one particle in each of $N$ wells, all in the single-particle ground state.

Proceeding with the length calculation in the usual way by substituting Equation \ref{eq:EntangledDensity} into Equation \ref{eq:LengthAgnostic} yields
\begin{equation}\label{eq:EntangledLength}
L^2 =  \frac {12} {N}  \sum_{i=1}^N \int_{-\infty}^{+\infty} dx \left( x^2 - \Braket{x}^2 \right) \rho_1^{(i)}(x) .
\end{equation}
The probability density can be expressed by a sum over the individual segments via
\begin{align}\label{eq:QuantumAdditivity}
\rho_1(x) = \frac {1} {N} \sum_{i=1}^N \rho_1^{(i)} (x) = \frac {1} {N} \sum_{i=1}^N \rho_1^{(1)} \left( x -(i-1)L_1 \right),
\end{align}
where the form of the second summation assumes that each segment is identical, but shifted by $i L_1$ to get the contribution from the $i^\text{th}$ one, i.e. $\rho_1^{(i)} \left( x \right) = \rho_1^{(1)} \left( x -(i-1)L_1 \right)$.  Since the probability density in one well can spill over into another one, Equation \ref{eq:QuantumAdditivity} effectively accounts for the contribution of particles in all wells to any given one.

This procedure is an implementation of the classical notion that the full length is the sum of its parts.  However, we leave $L_1$ arbitrary and solve for it self-consistently (below) demanding that the full length $L$ is given by $N L_1$.  This will test the validity of the classical notion of additivity assumed in Equations \ref{eq:PIB-Length-Cell} and \ref{eq:PIB-Length-Add-Cells}.

\begin{figure}[h]
\includegraphics{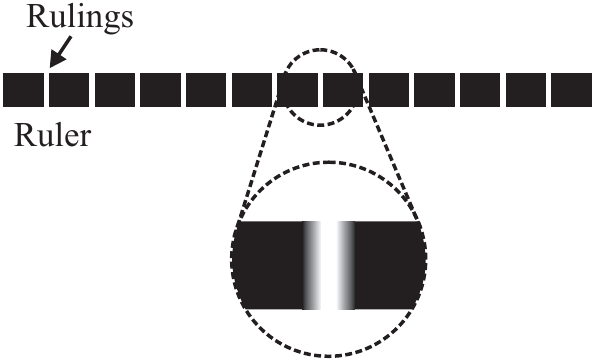}
\caption{The markings on a ruler become fuzzy on microscopic scales.}
\label{fig:FuzzyRuling}
\end{figure}
Consider the length formula given by Equation \ref{eq:SimplifiedL-Final}.  The term under the square root is the average of the length variance contributions from each of the occupied states.  This procedure makes sense for a monolithic object in which every electron is delocalized over the full length of the object.  Then, the probability density is the average of the contributions from each state.  This is precisely the kind of probability density that is depicted in Fig.~\ref{fig:FuzzyRuling}.

When the ruler is made by cutting a rod into segments, each segment localizes electrons.  The ground state configuration of the system is then given by each segment being in its ground state.
\begin{figure}[h]
\vskip 0in
\hspace{-.5em}\includegraphics[width=3.35in]{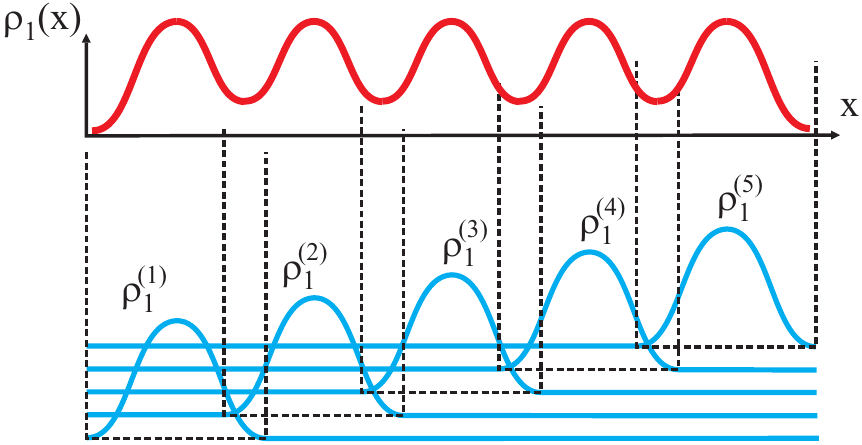}
\caption{The bottom curves show the probability density due to each segment, with each plot offset slightly upwards for readability.  The dashed lines show the infinite well walls and the solid curve at the top is a sum over each well's contribution that yields the system's probability density.}
\label{fig:Segments}
\end{figure}
Figure \ref{fig:Segments} shows a reconstruction of the full density when the segments are placed end-to-end with a spacing of $L_1$.  The density is then calculated using Equation \ref{eq:QuantumAdditivity}.  Figure \ref{fig:Segments} is in essence representing the probability density in the same way as in Fig.~\ref{fig:FuzzyRuling}, except the system is made of segments, with each contributing to the probability density in the local well as opposed to particles in different states of the same larger well.  To see that the total length is the average length per segment multiplied by the number of segments, we evaluate Equation \ref{eq:QuantumAdditivity} for the probability density in Fig.~\ref{fig:Segments}.

Some messy mathematics is required to solve for the quantum length wavefucntions spill over into adjacent wells.  These details are left to Appendix \ref{sec:MathDetails}.  Substituting Equation \ref{SumIi} into Equation \ref{eq:SegmentedLength2} yields the final result
\begin{align}\label{eq:SegmentedLengthFinal}
L^2 =  \left( \frac {1} {3} - \frac {1} {2 \pi^2} \right) a^2 + \left( N^2 - 1 \right) L_1^2 .
\end{align}

Recall that $L_1$ -- the length of the first segment -- needs to be determined.  We call upon translational invariance by a unit cell to demand that each segment is of the same length and that the full length is given by $L = N L_1$.  Applying this condition to Equation \ref{eq:SegmentedLengthFinal} yields
\begin{align}\label{eq:SegmentedSelfConsistentLength}
N^2 L_1^2 =  \left( \frac {1} {3} - \frac {1} {2 \pi^2} \right) a^2 + \left( N^2 - 1 \right) L_1^2 .
\end{align}
Solving Equation \ref{eq:SegmentedSelfConsistentLength} for $L_1$ yields
\begin{align}\label{eq:SelfConsistentLength}
L_1 = a \sqrt{\frac {1} {3} - \frac {1} {2 \pi^2} } .
\end{align}
Equation \ref{eq:SelfConsistentLength} is just the length of a single particle in a box in its ground state as given by Equation \ref{eq:BoxExcite-state-length}.  This shows that quantum parsing acts in the way we would expect from classical arguments based on additivity when stacking segments.

In conclusion, a ruler can be made by stacking individual pieces with spacing given by the quantum length of each piece.  While this type of parsing appears to be a classical construct, the same result is obtained through a self-consistent fully quantum calculation with the requirement of additivity as described above.  The quantum nature is evident by the length being shorter than the box size, but the classical act of stacking the segments is consistent with the quantum description.

\subsection{Bosons}

The same analysis can be performed for Bosons.  For the sake of argument, we assume that the bosons have zero spin so that each unit cell contributes one boson.  Equation \ref{eq:PIB-BosonLength} gives the length of a box filled with bosons in their ground state.  With $a = N a_0$, this gives a length of
\begin{equation}\label{eq:PIB-BosonLength-lattice}
L_N =   a_0 N \sqrt{  1 - \frac {6} { \pi^2 } } .
\end{equation}
If the box is chopped into $R$ pieces, the length of the individual pieces will be given by Equation \ref{eq:PIB-BosonLength} with $a = N a_0 / R$, or
\begin{equation}\label{eq:PIB-BosonLength-sections}
L_{N/R} =   \frac {a_0 N} {R} \sqrt{  1 - \frac {6} { \pi^2 } }
\end{equation}
and the total length is $L_R =R L_{N/R}$, given by
\begin{equation}\label{eq:PIB-BosonFull}
L_R =  a_0 N \sqrt{  1 - \frac {6} { \pi^2 } } .
\end{equation}
Thus for bosons, chopping the ruler into pieces and recombining them gives the same length, as is verified by the equality between Equations \ref{eq:PIB-BosonLength-lattice} and \ref{eq:PIB-BosonFull}.

Unlike the fermion case, where there is an ideal cutting interval, the boson ruler becomes more reliable as it is cut into more sections.  The length is always smaller than the size of the box that defines the potential.  Since the ruler is defined by the electron cloud, or in this case the boson cloud, this is the length that matters.

Finally, we calculate the ground state energy of the uniform ruler and one made by chopping it into sections.  The monolithic ruler energy for each particle is simply the energy of a particle in a box of length $N a_0$, so the total energy is $N$ times the single particle energy, yielding

\begin{align}\label{eq:GS-Boson-energy-FullRod}
E_G = \frac {h^2} {8 m N a_0^2}.
\end{align}
The ground state energy of each of the $R$ segments is of the form of Equation \ref{eq:GS-Boson-energy-FullRod} with $N \rightarrow N/R$, and the total energy is $R$ times the segment energy, yielding
\begin{align}\label{eq:GS-Boson-energy-Pieces}
E_G^\prime  = \frac {h^2 R^2} {8 m N a_0^2}.
\end{align}
Thus, the energy grows quadratically with the number of segments, requiring substantial work in producing each slice.  So, increased accuracy comes at a cost of increased energy.  Thus arbitrarily high precision is possible but comes at the cost of infinite energy.

Finally, we note that the boson system can be placed into a superposition with two particles per state to mimic the ground state fermion system, albeit with all positive signs in the superposition defining the entangled state. In this configuration, the signs are irrelevant, leading to the same results as obtained for fermions. Thus, bosons can be manipulated into many more configurations of energy equal to or less than the fermion ground state energy, leading to a richer set of configurations to investigate as an exercise for the student.

\subsection{Entanglement of Ruler and Object}

The spin statistics theorem demands that all electrons must be entangled, even if they do not interact.\cite{kuzyk19.01} Since matter is full of electrons, how does entanglement between the apparatus and the object being measured affect a measurement?  We can study this effect using the particle in a box model of a ruler and the object.

For $N$ noninteracting particles in a box of length $N a_0/2$, where  $a_0$ is the length of the unit cell that contains two particles, the energy of single-particle state $n$ is
\begin{equation}\label{eq:BoxEnergyNcells}
E_n = \frac {2 \pi^2 \hbar^2 n^2} { m a_0^2 N^2}.
\end{equation}
The Fermi energy for a box with an even number of electrons is calculated from Equation \ref{eq:BoxEnergyNcells} with $n = N/2$, yielding
\begin{equation}\label{eq:BoxFermiEnergyNcells}
E_f = \frac {\pi^2 \hbar^2 } {2 m a_0^2}.
\end{equation}
As such, the Fermi energy is independent of the size of the box since two electrons are added with each unit cell.  Figure \ref{fig:ObjectRuler} shows an object and a ruler that is made of the same material so the unit cells in each are the same size.  Since the ruler segment is smaller than the object, its single-particle ground state energy $E_0^r$ is higher than that of the object.  Furthermore, the energy levels of the object are more densely spaced than in the ruler.  We have arbitrarily chosen a ruler with 4 electrons and an object with 10 electrons for illustration.

Since the state vector of the 14 electron composite system is antisymmetric upon interchange of any two of them, the electrons in a universe made of only the ruler and the object are necessarily entangled, independent of the distance between them.  As such, we will express the state vector of the combined system as
\begin{align}\label{eq:Entangled}
\Ket{\Psi} & = \Ket{1,2,\dots,10,1^\prime, \dots 4^\prime} \nonumber \\
& = \frac {1} {\sqrt{14}} \left(\sum_{n=1}^{10} \Ket{n} \ket{\overline{n}} + \sum_{n^\prime =1}^4 \Ket{n^\prime} \ket{\overline{n}^\prime} \right) ,
\end{align}
where the primes indicate the state index of the ruler and the unprimed ones the state index of the object.  The barred index refers to the state that excludes that state and is the fully anisymmetrized one of all other state indices of both the ruler and the object.

Up to this point we have assumed that the object and ruler can be placed side by side and the measurement made by comparing the ruler and object with a scattering experiment, where the scattered light is measured by eye or with an instrument.  At ever smaller sizes, shorter wavelengths of the electromagnetic spectrum or particle beams of shorter de Broglie wavelength can be used as a probe.

With enough spatial resolution, a well-enough separated object and ruler appear distinct because the electron distribution is confined to one box or the other even though the electrons are entangled.  An image of the ruler is a mapping of the probability density $\left| \Braket{x^\prime|\Psi} \right|^2$, which is defined by
\begin{align}\label{eq:RulerProbDensity}
\rho(x^\prime) =\left| \Braket{x^\prime|\Psi} \right|^2 &= \left| \frac {1} {\sqrt{4}} \sum_{n^\prime =1}^4 \Braket{x^\prime | n^\prime} \Ket{ \overline{n}^\prime} \right|^2 \nonumber \\
& = \frac {1} {4} \sum_{n^\prime =1}^4 \left| \Braket{x^\prime | n^\prime } \right|^2 ,
\end{align}
where we use the fact that $\Braket{x^\prime | n} \Ket{\overline{n}} =0 $ because the primed position state vector acts only on the ruler state vectors.  Note that the probability density needs to be renormalized when calculating the length from this subsystem.  This primed part of the state vector is experimentally probed by illuminating only the ruler.  Thus, the probability density given by Equation \ref{eq:RulerProbDensity} will yield the length of the ruler.  The same process can be applied to the object, and the two compared.  Clearly, the ruler can be eliminated and the illuminated pixels of the camera used as the segments of a ruler to characterize the object's size.  Then, the electrons in the camera's pixels and the object are entangled.

\begin{table*}
\fbox{Case I} \hspace{23.5em} \fbox{Case II}
\centering
\begin{tabular}{|c|c||c|c|}
  \hline
  & & & \\[-0.7em]
  \hspace{.65em}\includegraphics{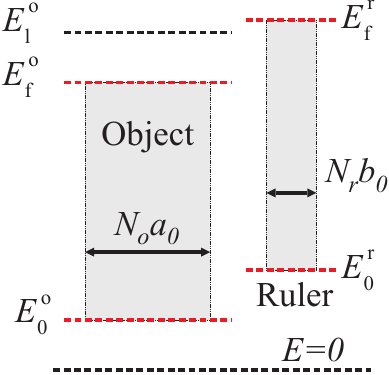}\hspace{.65em} & \hspace{.65em}\includegraphics{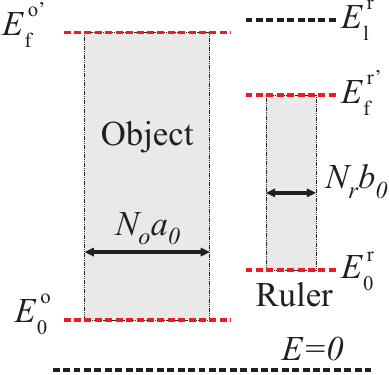}\hspace{.65em} & \hspace{.65em}\includegraphics{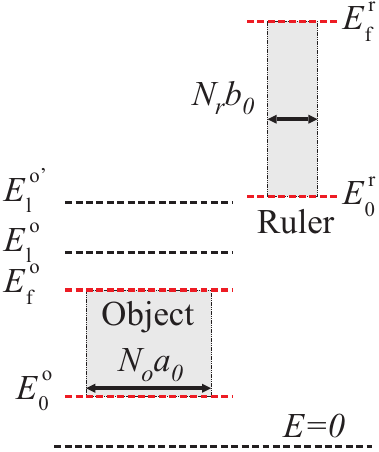}\hspace{.65em} & \hspace{.65em}\includegraphics{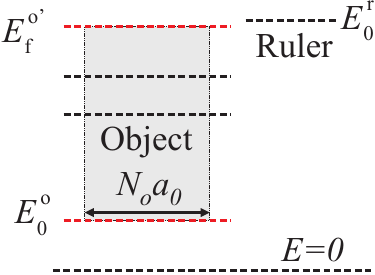}\hspace{.65em} \\[1em]
  $\frac {a_0} {b_0} > \frac {6} {5} $ &  $\frac {a_0} {b_0} > \frac {6} {5} $ & $\frac {a_0} {b_0}  > \frac {12} {5} $ & $\frac {a_0} {b_0} > \frac {12} {5} $ \\[1em]
  $\frac {N_o} {N_r} = \frac {10} {4} $ & $\frac {N_o^\prime} {N_r^\prime} = \frac {12} {2}$ & $\frac {N_o} {N_r} = \frac {10} {4} $ & $\frac {N_o^\prime} {N_r^\prime} = \frac {14} {0} $ \\[1em]
 Classical: $\frac {N_o a_0} {N_r b_0} = 3 $  & $\frac {N_o a_0} {N_r b_0} = 3$ & $\frac {N_o a_0} {N_r b_0} = 6$  & $\frac {N_o a_0} {N_r b_0} = 6 $ \\[1em]
Quantum:  $\frac {L_{10}^o} {L_4^r} = 3.45 $ & $\frac {L_{12}^o} {L_2^r} = 4.41 $  & $\frac {L_{10}^o} {L_4^r} = 6.91$ & $\frac {L_{14}^o} {L_0^r} = $ N.D. \\[1em]
(a) & (b) & (c) & (d) \\[1em]
  \hline
\end{tabular} \caption{Energy level diagram of an object and ruler segment.  The gray shaded areas represent the occupied states.  (a) A ruler with Fermi energy $E_f^r$ that is above the highest unoccupied state energy of the object, $E_l^o$, is not in its true ground state.  (b) The two electrons at the Fermi level of the ruler shown in Table \ref{tab:fermi}a de-excite into the lowest unoccupied state, leading to the true ground state of the system.  (c) All four electrons in the ruler segment are above the two lowest unoccupied states of the object.  (d) In the lowest energy configuration of the system shown in Table \ref{tab:fermi}c, all of the electrons leave the ruler, rendering it invisible.  The rows below the figures, show from top to bottom, the ratio of the lattice constant of the object and ruler $a_0/b_0$, the ratio of the number of lattice periods of the object to the number of lattice periods of the ruler $N_0/N_r$, the ratio of the classical lengths $N_0a_o/N_rb_0$, and the ratio of the quantum lengths calculated from Equation \ref{eq:PIB-Length}.  $n$ in the subscript of $L_n^o$ is the number of electrons in the object and $n$ in $L_n^r$ is the number of electrons in the ruling.  Note that the nuclei that define the potential are assumed to remain fixed so the lattice spacing does not change when an electron jumps from the ruler to the object.  In a real object, removal of electrons can make the system unstable.}  \label{tab:fermi}
\end{table*}

Things become more interesting when the ruler is made form a different material than the object, which is typical of real systems.  Then, the lattice spacing is different in each material.  Let the lattice spacings of the ruler and object be $b_0$ and $a_0$.  For the rest of the discussion, we assume that when the object and ruler are well separated, the object has 10 electrons and the ruler has 4 electrons as in Fig.~\ref{fig:ObjectRuler}.

When the lattice constants are different, the Fermi energies no longer line up, nor do the single-particle ground states.  The diagram in Table \ref{tab:fermi}a shows an example of the case when the unit cell of the ruler is smaller than that of the object.  The horizontal dashed lines show the lowest-energy single-particle state and the Fermi energy.  The shaded region represents all of the occupied states in between these two extremes.

The lowest unoccupied state energy of the object $E_l^o$ is given by Equation \ref{eq:BoxEnergyNcells} with $n=6$ and $N=10$, yielding
\begin{equation}\label{eq:ObjectLU}
E_l^o = \frac {18 \pi^2 \hbar^2 } {25 m a_0^2 }.
\end{equation}
The Fermi energy of the ruler $E_f^r$, given by Equation \ref{eq:BoxFermiEnergyNcells} with lattice spacing $b_0$ will match the highest unoccupied state energy $E_l^o$ given by Equation \ref{eq:ObjectLU} when
\begin{align}\label{eq:Fermi=HU}
E_f^r =  E_l^o \hspace{1em} \Rightarrow \hspace{1em} \frac {18 \pi^2 \hbar^2 } {25 m a_0^2 } = \frac {\pi^2 \hbar^2 } {2 m b_0^2}  ,
\end{align}
or
\begin{align}\label{eq:Fermi=HU-Critical}
\frac {a_0} {b_0} = \frac {6} {5} .
\end{align}
The energy-level diagram in Table \ref{tab:fermi}a is for a ratio that is slightly larger than $\frac {a_0} {b_0} = \frac {6} {5} $ .

When $b_0 < 5 a_0 /6$, the Fermi energy of the ruler is higher than the lowest-energy unoccupied state of the object so the entangled ruler/object system is not in its ground state.  If the electrons are noninteracting as assumed from the start, the system will remain in this excited state.  However, for a deep but non-infinite well, the two electrons at the Fermi level of the ruler will tunnel to the object.  As a result, the object will become longer and the ruler will become shorter, as depicted in Table \ref{tab:fermi}b.  If the nuclei, which define the potential remain fixed, the lattice constant and therefore the well size stays the same.

The numbers in Table \ref{tab:fermi} summarize the results that follow.  $N_o$ and $N_r$ are the number of unit cells that form the object and the ruler.  The classical size is given by the number of unit cells times the unit cell size, which is given by the width of the box.  The ruler determines the length of the object by counting the number of segments placed end-to-end across the object.  In the classical case, this ratio is $\frac {N_o a_0} {N_r b_0} = 3 $.  Thus, we would say that the object size is three ruler segments.

The quantum length of the object before the electrons transfer to the object is given by Equation \ref{eq:PIB-Length} with $N=10$ and lattice constant $a_0$, and the ruler length is given by Equation \ref{eq:PIB-Length} with $N=4$ and lattice constant $b_0$  yielding the ratio $L_{10}^o / L_4^r = 3.45$.  In the quantum case, the object is found to be 3.45 units in length.

After the system relaxes into its ground state, the object has 12 electrons and the ruler has 2, as shown in Table \ref{tab:fermi}b.  As a result, the object's Fermi energy has increased and the ruler's Fermi energy has decreased.  The quantum lengths of the ruler and the object have both changed, resulting in the conclusion that the object's length is 4.41 units.  Thus, there are several quantum effects at play.  First, the quantum length and classical length are different.  Secondly, the length determined from a standard measurement is affected by the fact that the electron and ruler are not independent systems, and that they share electrons.  As a result, both the object and the ruler can change length.

In doing a measurement, it is important to know which configuration is being characterized.  Is it the one in which the ruler and the object are independent, or the one in which electrons are exchanged?  The process of electron transfer can in principle be established with the detection of emitted photons.  Each electron jump will be associated with the emission of a photon with energy $\hbar \omega$ given by the difference between the Fermi energies, or $\hbar \omega = E_f^r - E_f^0$.  The emission of a single photon of this energy is a signal that only one electron is exchanged, so the ruler and object would then have an odd number of electrons, yielding yet a different measure of the length.

An even more peculiar configuration is one in which all of the energy levels of the ruler are above vacant states of the object.  Table \ref{tab:fermi}c show the case where the lowest-energy occupied state of the ruler is above the lowest unoccupied state of the object, or $E_0^r > E_l^o$; and, the ruler's Fermi energy is above the second unoccupied state of the object, or $E_f^r > E_l^{o^\prime}$. The classical length of the object is 6 ruler units before the electrons are exchanged, and the quantum length is 6.91 ruler units.  However, once the electrons are exchanged, the ruler disappears because there are no longer any electrons in that segment.

In a real system, the material is held together by bonds that are mediated by interactions between the electrons and the nuclei.  If all electrons were to leave the ruler, it would fly apart.  The configurations of real materials do not have the required energetics for this to happen, nor are the lattice constants independent of the number of electrons.  However, the model systems presented here do not violate any quantum principles, illustrating the bizarre consequences of measuring length and the classical notion that the length is an additive inviolable property.

Other examples can be contrived.  For example, electrons can be made to transfer to the ruler from the object, leading to the evaporation of the object in the process of its length being measured.  In the quantum realm, there is no distinction between the ruler and the object, and the two remain entangled through the electrons from which they are made and through which we detect them.

\subsection{Light Ruler}\label{sec:lightRuler}

This work has focused on size in the non-relativistic quantum realm.  As such, the length is the one determined in the rest frame of the object, and, the electrons' motions are slow compared with the speed of light.  As the ruler and object gets ever smaller, relativistic effects would result in a slew of phenomena that will not be discussed here.  However, the idea of using a light ruler as is commonly applied to determining length and time in relativistic thought experiments deserves comment.

A light ruler uses a burst of light that is emitted from one end of an object, and reflected from a mirror at its other end.  The round trip time $T$ is measured and the length determined using the speed of light $c$ via $L = ct/2$.  One can imagine other designs using parallel mirrors to make an interferometer and counting fringes, etc.  In all such scenarios, the basis of the length measurement is reflection form a mirror.

A length measurement requires that a mirror be attached to one end of the object.  This leads to various complications that depend on the how the mirror is attached.  Instead, imagine that the object is transparent and that the light pulse propagates through the material and reflects off of its end.  For a large object with many electrons, the ends of the object are sharp, so the light reflects in the usual way.  As the object is made smaller, the edge becomes fuzzy.  If the width of the fuzziness is greater than the wavelength of the light, the light will no longer reflect.

To increase the accuracy of the measurement, the light pulse needs to be made shorter, but that requires a broader range of wavelengths. Since only those wavelengths that are longer than the fuzziness of the edge get reflected, the reflected pulse will be much broader than the incident one, leading to a decrease in the accuracy.  Thus, the light ruler will be constrained to the same limitations that we found using ruler segments.  This exercise would be a useful pedagogical one, and deserves future attention.

\section{Lessons Learned}\label{sec:Learned}

We started with classical reasoning to argue that the length of a spatially-localized object with sharp edges is given by the spread of the object's density as quantified by the position variance and average, which gives the expected answer.  Then by extrapolation we made the ansatz that in the quantum realm, the length is given by the position uncertainty and found that the length so defined becomes the classical length in the many-electron limit of particles in a box.  This illustrates to the student how models are proposed and tested.

This exercise suggests that it is possible to form a macroscopic object that is a purely quantum in nature.  A macroscopically large box with many noninteracting electrons is such a system.  However, any blemish in the material, no matter how minute, localizes electrons.  As a consequence, real materials that have many imperfections will act as a quantum system only in those regions between imperfections.  Furthermore, a system that interacts with the environment de-coheres, washing out any vestiges of a quantum signature.  As such, making a macroscopic quantum system for the purposes of studying properties such as length is untenable; but, thought experiments pondering quantum length on macroscopic scales provide useful insights.  Such reasoning teaches students about the correspondence principle and when quantum mechanics breaks down in the real macroscopic world.

Since the quantum nature of length is suppressed at length scales that are visible to the human senses, a material's length is given by the sum of the lengths of its constituents, which can also be defined by a quantum length.  This shows how our intuition of length additivity is based on localized sections of a ruler, and need not be generally true.

The ground state of a boson ruler, on the other hand, does not approach uniform density with sharp edges in the classical limit; but, it is possible to construct an excited state made of a particular superposition that meets the criteria of uniformity with sharp boundaries.  There is no reason to believe that our experience with ordinary matter, whose properties are dominated by fermions and the statistics that they follow, would inform us of how a ruler made of bosons would act.  This teaches us how our intuition is based on special cases and why other systems or realms appear so bizarre.

Classical reasoning would lead us to conclude that rulers become more precise when the gradations are made smaller.  This is so until quantum effects become important, at which point the length of the ruler made of slices is shorter than the original monolithic one.  Thus, the undivided length of the ruler is no longer the sum of the lengths of its parts; so, applying additivity to the slices yields measures of an object's length that depends on the size of the rulings.

In such finely graded rules, quantum corrections need to be taken into account to get the object's true length.  This interplay between increasing resolution of the ruler and the onset of nonadditivity of the classical quantities leads to an optimum number of ruler parsings, equal to the square root of the number of unit cells from which the ruler is constructed.  Since slicing the ruler requires work, the increased accuracy comes at an energy cost.  This illustrates the relationship between energy and accuracy, and is analogous to the application of Heisenberg's uncertainty principles to understand why higher-energy particle accelerators are required to probe smaller length scales.

There are no unexpected consequences arising from the entanglement of electrons that are shared by the ruler and the object to be measured  when they are made of the same material.  Then, the occupied states in the ruler and in the object are no different than if the two were unentangled objects, thus agreeing with the classical picture of distinct objects.  However, if the two are made of different materials with differing unit cell sizes, the Fermi energies no longer match.  When the size of the ruler segment is small enough to bring its Fermi energy above the energy of the lowest unoccupied state of the object, an electron can jump from the ruler to the object, affecting the length of each, changing the nature of the measurement and making the result ill-defined.  In the extreme case of large unit-cell size mismatch, all of the electrons can jump from the ruler to the object, in essence causing it to disappear since the electrons with which light interacts are absent.  The ruler and object can be reversed to make electrons jump to the ruler, leading to similar weirdness.

\section{Conclusion}

Perhaps one of the most useful consequences of going through the exercise of defining a quantum length is the fresh perspective that it might bring to topics beyond length itself.

For example, what lengths do other potentials give?  The harmonic oscillator's particle density expands when excited to higher state energy, so the calculated length grows with energy rather than converging to a constant value as for the box.  In such cases, the classical turning points might be a more reasonable characterization of a length scale, which is not the length but rather a range of motion.  Interestingly, for a particle in a box, the turning points and the classical length are the same.  For a wave packet, which describes the classical particle attached to a spring, a calculation of its length will give the width of the wave packet, thus the size of the particle rather than the size of the well.  Alternatively, filling the harmonic well with many particles leads to a flattening of the probability density and the wiggles in the wave function disappear.  This has a classical feel.  Unfortunately, the many-particle harmonic oscillator ground state does not have a classical analog so does not add new insights into the classical/quantum transition.  Such considerations make the student think about the general applicability of the concepts.

Length brings up broader issues.  Central to the classical definition of length is one-to-one correspondence\cite{woods20.01} between the segments of a ruler and points on the object.  Such correspondence between the two assumes that numbers are agnostic with regards to the composition of the objects that they are quantifying.\cite{dantz05.01,ifrah00.01,flegg02.01,flegg89.01} In the quantum realm, the composition of the material can affect the measurement of its length.  Furthermore, on small enough scales, objects are no longer distinct.  This suggests that one-to-one correspondence is a classical approximation that holds on human size scales, but breaks down when an object is made of only a few electrons.

Counting is the basis of mathematics, so it is remarkable that the quantum realm -- where one-to-one correspondence is ill-defined -- is so well described by a theory based on counting.  Size is thus a useful focus for studying correspondence and when it fails in this role.  Most importantly to the student, this paper illustrates how defining a general but simple concept of length reveals weaknesses in classical concepts that in the end builds a deeper intuition about the transition from the classical to quantum realm.

\vspace{1em}

\noindent{\bf Acknowledgements} I thank the National Science Foundation (EFRI-ODISSEI:1332271 and ECCS-1128076) and the Meyer Distinguished Professorship of the Sciences for generously supporting this work.  I am indebted to the reviewers for their substantive input, which greatly improved this manuscript.

\appendix

\section{Varying Densities}\label{sec:NonuniformRod}

From the perspective of classical physics, macroscopic objects are made from infinitely divisible parts.  The underlying assumption is that each segment, no matter how small, retains the properties of the material.  Then, bulk properties of a material arise from a sum over its infinitesimal parts.

For example, the length of a curve is determined by breaking it down into small sections that are each assumed to be line segments, and summing over the lengths of the individual sections.  In the limit of infinitesimal segments, the sum becomes an integral.

Because the length of a line between the endpoints given by coordinates ${\bf r}_i$ and ${\bf r}_f$ is given by
\begin{align}\label{eq:Pythagoras}
L = \sqrt{(x_f - x_i)^2  + (y_f - y_i)^2 + (z_f - z_i)^2} ,
\end{align}
then the length of a curve between those endpoints is given by
\begin{align}\label{eq:Pythagoras}
L = \int_{{\bf r}_i}^{{\bf r}_f} \sqrt{dx^2  + dy^2 + dz^2} ,
\end{align}
which can be calculated from the parametric form of the curve ${\bf r} (t) = (x(t), y(t), z(t))$.

In analogy, Equation \ref{eq:ClassicalLength} holds for a uniform material, so the length of a nonuniform rod is determined from the sum over the lengths of the segments that are each small enough for them to be uniform, yielding
\begin{equation}\label{eq:LengthAgnosticIntegrate}
L = \sum_i \sqrt{12  \int_{x_i}^{x_{i+1}} \rho_i \left(  x^2 - \bar{x}^2 \right) dx, }
\end{equation}
where $\rho_i$ is defined by Eq.~\ref{eq:ClassicalPsi} to be
\begin{equation}\label{eq:ClassicalSegLen}
\rho_i = \left| \psi \right|^2 = \frac {1} {x_{i+1} - x_i} .
\end{equation}
The spacing $x_{i+1} - x_i$ can be chosen to be different for each $i$ as long as the segments are uniform.

To check if Equation \ref{eq:LengthAgnosticIntegrate} gives the expected length of a nonuniform rod, we focus on the argument of the square root in Equation \ref{eq:LengthAgnosticIntegrate}, which defines the length of segment $i$ to be
\begin{equation}\label{eq:LengthAgnosticLi}
L_i =  \sqrt{ 12  \int_{x_i}^{x_{i+1}} \frac {  x^2 - \bar{x}^2 } {x_{i+1} - x_i} dx, }
\end{equation}
where we use Equation \ref{eq:ClassicalSegLen} for the normalized density.  With $\bar{x}$ being the average position on the interval $[x_{i+1} , x_i]$
\begin{equation}\label{eq:AveLength}
\bar{x} = \frac {x_{i+1} + x_i} {2} ,
\end{equation}
integration of Equation \ref{eq:LengthAgnosticLi} yields
\begin{equation}\label{eq:Li}
L_i = \left| x_{i+1} - x_i \right| .
\end{equation}
The student should do the integration to verify Equation \ref{eq:Li}.

Equation \ref{eq:Li} shows that the classical length of the rod is simply the sum over the segments, as we would expect.  Note that segments of nonzero density add to the length and those of non-zero density do not.  Thus, a rod with a gap will give a length that excludes the gap.

\section{Other Length Definitions}\label{sec:Third-Order-Length}

There are many ways to construct a linear combination of expectation values using the various functions of the length that meet the criteria for the length definition.  We might be tempted to start with the next less simple expression by considering the quantities $\braket{ x^3}$, $\braket{ x^2} \braket{x}$ and $\braket{x}^3$.  But if the density is uniform, all three of these quantities vanish, as the reader can verify simply by noticing that the expectation values are integrals over odd functions of $x$ so must vanish.

The next case is are expectation values that give the fourth power of length.  It is left as an exercise for the student to determine the linear combination of these expectation values that is translationally invariant and gives the correct length for a uniform rod.  The result is
\begin{align}\label{eq:quarticLength}
L_4 =  2\sqrt[3]{ 5 \left( \braket{ x^4} - 2 \braket{ x^3} \braket{x} + \braket{x}^4 \right) }.
\end{align}
The subscript ``4" refers to the length being defined in terms of fourth moments.  The equation used in the main text is the simplest length given by $L_2$.

The reader can easily verify that Equation \ref{eq:quarticLength} gives the length $L$ using Eqs.~\ref{eq:<x>}, \ref{eq:<x^2>},
\begin{align}\label{eq:<x^4>}
<x^3> = \frac {L^3} {4}
\end{align}
and
\begin{align}\label{eq:<x^4>}
<x^4> = \frac {L^4} {5} ,
\end{align}
for a rod between the origin and $x=L$.

\begin{figure}[h]\label{fig:L2-L4-Compare}
\vskip 0in
\hspace{-.5em}\includegraphics{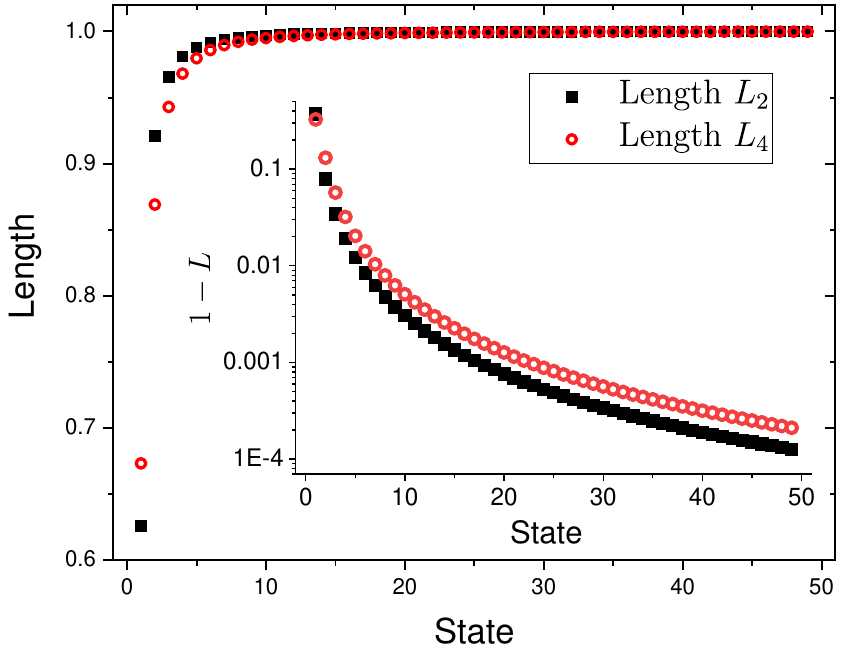}
\caption{The lengths $L_2$ and $L_4$ as a function of state index for a quantum particle in a box.  The inset shows a log plot of the deviation of the length from the classical value.}

\end{figure}

Figure \ref{fig:L2-L4-Compare} compares the length as a function of state index for a quantum particle in a box for the two different forms for the length.  Equation \ref{eq:quarticLength} gives a value of $L_4$ that is closer to the classical value than $L_2$.  However, $L_2$ converges more rapidly to the classical length as shown in the log plot in the inset.

\section{Many-Particle Systems}\label{sec:ManyParticleNotate}

\subsection{Wave Functions}

This appendix introduces a notation that both reduces confusion and simplifies taking expectation values of operators, using two and three particles for illustration.  The reader who is interested in the general results can skip to the next appendix.

For two identical particles, being either fermions or bosons, the state vector is expressed as\cite{duck98.01}
\begin{equation}\label{DefineKet}
\Ket{n,m} = \frac {1} {\sqrt{2}} \Big( \Ket{n} \Ket{m} \pm \Ket{m} \Ket{n} \Big) ,
\end{equation}
where the plus and minus signs are for bosons and fermions.  We understand $\Ket{n} \Ket{m}$ to mean  ``particle \#1 is in state $n$ and Particle \#2 is in state $m$."  Thus, when kets are written side-by-side, from left to right, they represent Particle \#1, Particle \#2, and so on.  Even though the particles are indistinguishable, they are not treated as such at this point.

The ket $\Ket{n,m}$ is expressed in agnostic form because it represents two particles that occupy states $n$ and $m$ without distinguishing which one is in which state.  If we exchange the two particles we find that $\ket{n,m} \rightarrow \pm \ket{m,n}$.  Since state vectors are unique up to a phase, both of them represent the same state, so we chose to express the arguments in ascending order of their numerical values, i.e. $n<m$.

We have used the classical concept of distinguishability to build a quantum-mechanical object that no longer distinguishes between its parts.  We started by treating the particles as distinguishable, wrote an expression with built-in indistinguishability, then got a purely quantum object as embodied by Equation \ref{DefineKet}.

It follows that for a three-particle system, the wave function is given by
\begin{eqnarray}\label{Define3Ket}
\Ket{n,m,\ell} &=& \frac {1} {\sqrt{6}} \Big(\Ket{n} \Ket{m} \Ket{\ell}  \pm \Ket{n} \Ket{\ell} \Ket{m} + \Ket{\ell} \Ket{n} \Ket{m} \nonumber \\
&\pm& \Ket{l} \Ket{m} \Ket{n} + \Ket{m} \Ket{\ell} \Ket{n} \pm \Ket{m} \Ket{n} \Ket{\ell} \Big) .
\end{eqnarray}
We can express Equation \ref{Define3Ket} in a form that singles-out the state vector of Particle \#1
\begin{equation}\label{Define3KetContract1}
\Ket{n,m,\ell} = \frac {1} {\sqrt{3}} \Big(\Ket{n} \Ket{\bar{n}}  + \Ket{m} \Ket{\bar{m}} + \Ket{\ell} \Ket{\bar{\ell}} \Big) ,
\end{equation}
where
\begin{equation}\label{Define3KetContract2}
\Ket{\bar{n}} = \frac {1} {\sqrt{2}} \Big( \Ket{m} \Ket{\ell} \pm \Ket{\ell} \Ket{m} \Big),
\end{equation}
and where the other ``barred" states are obtained by comparing Equations \ref{Define3KetContract1} and \ref{Define3Ket}.  Thus, we can view $\Ket{\bar{n}}$ as the state that remains when Particle \#1 is removed.  Note that Equation \ref{Define3KetContract1} shows all plus signs because the signs can be absorbed into the barred states.  It is straightforward to verify that the barred states are orthonormal, or
\begin{equation}\label{BarOrthoNorm}\Braket{\bar{n}| \bar{n}^\prime} = \delta_{n,n^\prime}.
\end{equation}

The contracted form given by Equation \ref{Define3KetContract1} is useful in situations where the property of Particle \#1 is to be calculated.  For example, the probability density of Particle \#1 is given by
\begin{eqnarray}\label{Define3Density}
&& \rho(x_1) = \left|\Braket{x_1|n,m,\ell} \right|^2 \nonumber \\
&=& \left| \frac {1} {\sqrt{3}} \Big( \Braket{x_1 | n} \Ket{\bar{n}} + \Braket{x_1 | m} \Ket{\bar{m}} + \Braket{x_1 | \ell} \Ket{\bar{\ell}} \Big) \right|^2 . \nonumber \\
\end{eqnarray}
Orthonormality of the barred states gives
\begin{equation}\label{Desnity3Final}
\rho (x_1) = \frac {1} {3} \Big( \rho_n (x_1) + \rho_{m} (x_1) + \rho_\ell (x_1) \Big),
\end{equation}
where $\rho_i (x_1)$ is the probability density of particle \#1 in state $i$.  Note that the sign of the barred state is irrelevant because they always appear in pairs of the form $\Braket{\bar{n}|\bar{n}}$.  Since the particles are indistinguishable, the result can be no different if we had done the calculation for any other particle.  Thus, Equation \ref{Desnity3Final} is the probability density for the system of three particles as a whole, and can be expressed simply as $\rho(x)$.

\section{Many-Particle States}

All of the above concepts apply to many-particle states.  For fermions, as is evident from the antisymmetric nature of the wave functions, if two particles occupy the same state, the wave function vanishes, as required by the Pauli exclusion principle.  As such, the ground state $\Ket{G}$ of the $N$-particle fermionic system can be expressed as the determinant
\begin{equation}\label{slater}
\Ket{G} = \frac {1} {\sqrt{N!}} \left|
  \begin{array}{cccc}
    \left| 1 \right> &  \left| 2 \right> & \dots & \left| N \right> \\
    \left| 1 \right> &  \left| 2 \right> & \dots & \left| N \right>  \\
    \vdots & \vdots & \ddots & \vdots \\
    \left| 1 \right> &  \left| 2 \right> & \dots & \left| N \right>  \\
  \end{array}
\right| ,
\end{equation}
where we label the single-particle ground state $\Ket{1}$.  Note that we use uppercase symbols to describe many particle states such as $\Ket{G}$, and lowercase symbols for one-particle states, such as $\Ket{n}$.

Similarly, any state of noninteracting fermions can be represented by
\begin{equation}\label{slater2}
\Ket{n_1, n_2, \dots, n_N} = \frac {1} {\sqrt{N!}} \left|
  \begin{array}{cccc}
    \left| n_1 \right> &  \left| n_2 \right> & \dots & \left| N \right> \\
    \left| n_1 \right> &  \left| n_2 \right> & \dots & \left| N \right>  \\
    \vdots & \vdots & \ddots & \vdots \\
    \left| n_1 \right> &  \left| n_2 \right> & \dots & \left| N \right>  \\
  \end{array}
\right| ,
\end{equation}
or by superpositions of states of the form given by Equation \ref{slater2}.  The contracted form given by Equation \ref{Define3KetContract1} is made clear by recognizing that the barred state vector $\Ket{\bar{n}}$ is just the determinant of the $(N-1) \times (N-1)$ matrix formed by excluding the $n^{\text{th}}$ row and $n^{\text{th}}$ column of the matrix.  Then, we can write the determinant in the more compact form
\begin{align}\label{eq:GeneralState}
\Ket{n_1, n_2, \dots, n_N} = \frac {1} {\sqrt{N}} \sum_{i=1}^N \Ket{n_i} \Ket{\bar{n}_i}
\end{align}

Note that it is simpler to write Equation \ref{slater} than Equation \ref{slater2}, so we can express state vectors in the form of Equation \ref{slater} and get the more general result by making the substitution $\Ket{i} \rightarrow \Ket{n_i}$.  If any two indices are the same, i.e. $n_i = n_j$, two of the columns will be the same and the determinant vanishes.  This enforces the Pauli exclusion principle -- the state vector vanishes if any pair of fermions are in the same state.  For fermions, then, Pauli exclusion is enforced by constructing state vectors in which all occupied single-particle states are different.

\subsection{Many-Particle Matrix Elements}

To get the expectation of an operator $A_1$, which acts only on Particle \#1, we use Equation \ref{eq:GeneralState}, which yields
\begin{align}\label{ExpectNFinal}
& \Braket{n_1, \dots n_N | A_1 | n_1, \dots n_N} \nonumber \\
&= \frac {1} {\sqrt{N}} \sum_{i=1}^N \Bra{n_i} \Bra{\bar{n}_i} A_1  \frac {1} {\sqrt{N}} \sum_{i^\prime=1}^N \Ket{n_{i^\prime}} \Ket{\bar{n}_{i^\prime}} \nonumber \\
&=  \frac {1} {N} \sum_{i=1}^N \sum_{i^\prime=1}^N \Braket{n_i| A_i | n_{i^\prime} } \Braket{\bar{n}_i | \bar{n}_{i^\prime}},
\end{align}
where in the last equality we have used the fact that $A_1$ operates only on the first ket.  Using orthonormality of the barred states, Equation \ref{ExpectNFinal} reduces to
\begin{align}\label{ExpectManyFinal}
\Braket{n_1, \dots n_N | A_1 |n_1, \dots n_N} = \frac {1} {N} \sum_{i=1}^N \Braket{n_i|A_1|n_i}  .
\end{align}

The notation used above is a bit sloppy.  The operator $A_i$, which acts on Particle \#i, should more rigourously be expressed as
\begin{equation}\label{SinceParticleOp}
A_i \rightarrow  \prod_{j \neq i}  \mathbb{1}_j A_i ,
\end{equation}
where $\mathbb{1}_j$ is the identity operator in the Hilbert space of particle $j$. So for example, we can define the many-particle position operator in terms of the single particle operators using
\begin{equation}\label{PositionOperator}
X = \sum_{i=1}^{N} \prod_{j \neq i}  \mathbb{1}_j x_i ,
\end{equation}
which is often imprecisely written as
\begin{equation}\label{PositionOperatorImprecise}
X = \sum_{i=1}^{N} x_i,
\end{equation}
where the identity operators are implicitly understood.

Calculations for electrons must include the spin, and account for the Pauli exclusion principle to fill the single-particle states.  To do so, we simply define the ground state vector of a spin-up electron to be $\left| 1 \right>$ and a spin down-electron to be $\left| 2 \right>$; the first excited statevector of a spin-up electron and spin-down electron are designated $\left| 3 \right>$ and $\left| 4 \right>$; and so on.  Thus, an odd integer $n$ represents a spin-up electron in state $\left| \frac {n+1} {2}, + \right>$ and an even integer $n$ represents a spin-down electron in state $\left| \frac {n} {2}, - \right>$.  In this latter representation, the first argument of the ket labels the single particle energy eigenstate and the second argument labels the spin.  The former notation with only one index is the simplest, especially when sums are involved, but the latter notation is more straightforward when evaluating sums over occupied states.

\section{Mathematical Details Overlapping Wells}\label{sec:MathDetails}

\begin{figure}[h]
\vskip 0in
\includegraphics{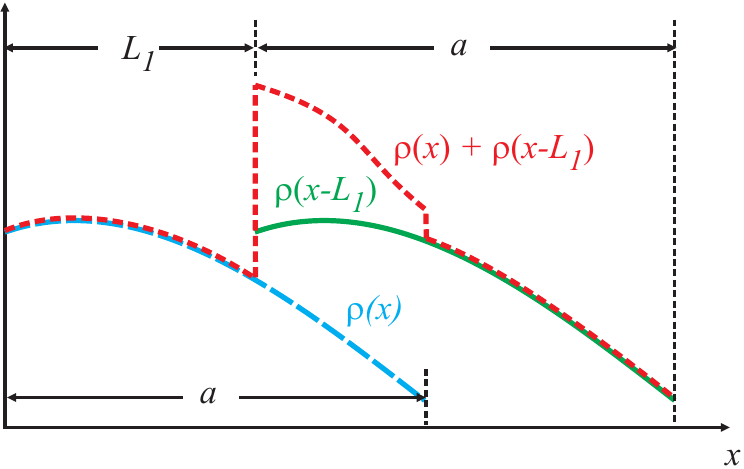}
\caption{The function $\rho(x)$ (long dashed curve), the shifted function $\rho(x-L_1)$ (solid curve), and the sum of the two (short dashed curve).}
\label{fig:FunctionSplice}
\end{figure}

This appendix calculates the expectation of a function $f(x)$ for a probability density that is made of a sum of functions shifted by $L_1$ as shown in Fig.~\ref{fig:FunctionSplice}.  The expectation values can be separated into integrals of three separate regions as follows:
\begin{align}\label{Eq:EvalauteShifted}
\Braket{f(x)} &= \int_0^{L_1} f(x) \rho(x) \, dx \nonumber \\
&+ \int_{L_1}^a f(x) \left( \rho(x) + \rho(x-L_1) \right) \, dx \nonumber \\
&+ \int_a^{L_1+a} f(x) \rho(x-L_1) \, dx ,
\end{align}
where we have used the fact that the probability density $\rho(x)$ is non-vanishing only in the interval $a$ of its own unit cell. Combining the first line of Equation \ref{Eq:EvalauteShifted} with the first term in parentheses of the second line and grouping the second term in parentheses of the second line with the third yields
\begin{align}\label{Eq:FinalShifted}
\Braket{f(x)} &= \int_0^a f(x) \rho(x) \, dx + \int_{L_1}^{L_1 + a} f(x) \rho(x-L_1) \, dx .
\end{align}

We can now evaluate the length using Equation \ref{eq:LengthAgnostic} by casting Equation \ref{eq:QuantumAdditivity} in the form of Equation \ref{Eq:FinalShifted} with $f(x) = x^2 - \Braket{x}^2$, which yields the square of the length
\begin{equation}\label{eq:SegmentedLength}
L^2 =  \frac {12} {N}  \sum_{i=1}^N \int_{(i-1)L_1}^{(i-1)L_1+a} dx \, \left( x^2 - \Braket{x}^2 \right) \rho_1^{(1)}\big(x-(i-1)L_1 \big) .
\end{equation}
Because the well is centrosymmetric, the expectation $\Braket{x}$ of a segment is just the position of its center.  For the $i^{\text{th}}$ segment, the center position is given by $\Braket{x}_i = L_1 (i-1) + a/2$.   $\Braket{x}$ is thus given by
\begin{align}\label{eq:SegmentedPosition}
\Braket{x} &= \frac {1} {N}  \sum_{i=1}^N \int_{(i-1)L_1}^{(i-1)L_1+a} dx \, x \rho_1^{(1)}\big(x-(i-1)L_1 \big) \nonumber \\
&= \frac {1} {N}  \sum_{i=1}^N \Braket{x}_i = \frac {1} {N}  \sum_{i=1}^N \Big( L_1 (i-1) + a/2 \Big).
\end{align}
Evaluating the sum yields
\begin{align}\label{eq:FullAveragePosition}
\Braket{x} = \frac {(N-1)L_1 + a} {2} .
\end{align}
Equation \ref{eq:SegmentedLength} then becomes,
\begin{align}\label{eq:SegmentedLength2}
L^2 &=  \frac {12} {N}  \sum_{i=1}^N \int_{(i-1)L_1}^{(i-1)L_1+a} dx \, x^2  \rho_1^{(1)}\big(x-(i-1)L_1 \big) \nonumber \\
&- 3 \big( (N-1)L_1 + a \big)^2 \nonumber \\
& \equiv  \frac {12} {N}  \sum_{i=1}^N I_i - 3 \big( (N-1)L_1 + a \big)^2 .
\end{align}

To evaluate the integral $I_i$ in Equation \ref{eq:SegmentedLength2}, we make the substitution $y = x - (i-1)L_1$, yielding
\begin{align}\label{eq:QuantumSegmentInegral}
I_i &=  \int_0^{a} dy \left( y + (i-1) L_1 \right)^2 \rho_1^{(1)} \left(y \right) \nonumber \\
&= \int_0^{a} dy \left( y^2 + 2y (i-1) L_1 + (i-1)^2 L_1^2 \right) \rho_1^{(1)} \left(y \right) .
\end{align}
The first term in Equation \ref{eq:QuantumSegmentInegral} is evaluated using Equation \ref{eq:x^2BoxMatrix} with $n=1$.  The second term can be evaluated using $\Braket{y} = a/2$ and the last term by noting that an integral over the probability density is unity by virtue of normalization.  Putting it all together, Equation \ref{eq:QuantumSegmentInegral} becomes
\begin{align}\label{eq:QuantumSegmentInegralEval}
I_i &= a^2 \left( \frac {1} {3} - \frac {1} {2 \pi^2} \right) + a (i-1)L_1 + (i-1)^2 L_1^2 .
\end{align}

Next we group the terms in Equation \ref{eq:QuantumSegmentInegralEval} so as to yield a polynomial in $i$, yielding
\begin{align}\label{eq:InegralRegroup}
I_i = \left[L_1^2 \right] i^2 &+ \left[ a L_1 -2 L_1^2 \right]i \nonumber \\
&+ \left[ a^2 \left( \frac {1} {3} - \frac {1} {2 \pi^2} \right) + L_1^2 - aL_1 \right],
\end{align}
where the terms in brackets are the coefficients.  Using $\sum_{i=1}^N 1 = N$, $\sum_{i=1}^N i = N(N+1)/2$ and
\begin{align}\label{n2sum}
\sum_{i=1}^N i^2 = N(N+1)(2N+1)/6 ,
\end{align}
after some algebra, we get
\begin{align}\label{SumIi}
\frac {1} {N} \sum_{i=1}^N I_i &=   \left( \frac {1} {3} - \frac {1} {2 \pi^2} \right) a^2 + \frac {N-1} {2} aL_1 \nonumber \\
&+ \frac { (2N - 1)(N-1)} {6} L_1^2 .
\end{align}




\end{document}